\documentstyle[12pt,aaspp4,epsf]{article} 
\def\undersim#1{\mathop{\vtop{\ialign{##\crcr
     $\hfil\displaystyle{#1}\hfil$\crcr\noalign
     {\kern1pt\nointerlineskip}\hbox{$\hfil\sim\hfil$}\crcr
     \noalign{\kern1pt}}}}}
\newcommand{\mettips}[8]
{
\noindent
\begin{figure}[t]
  \begin{center}  
      \leavevmode
      \epsffile{#8}
  \end{center}
    \caption[\protect\small \it #6]{\protect\small \it  #7 \label{#1}}
\end{figure}
}
\begin{document}

\title{A comparison between Fast Multipole Algorithm and Tree--Code to
evaluate gravitational forces in 3-D}

\author{R. Capuzzo--Dolcetta }
\affil{Istituto Astronomico, Universit\`a ``La Sapienza'', Rome, Italy}

\and

\author{P. Miocchi }
\affil{Dipartimento di Fisica, Universit\`a ``La Sapienza''  Rome, Italy}
\vfill

\vfill\eject
\vfill
\centerline{running head: FMA and Tree--Code comparison}
send proofs to: \centerline{R. Capuzzo--Dolcetta}
\par\centerline{Istituto Astronomico, Universit\'a La Sapienza}
\par\centerline{via G.M. Lancisi 29, I-00161, Roma, Italy}
\par\centerline{e--mail: dolcetta@astrmb.rm.astro.it}


\begin{abstract}
We present tests of comparison between our versions of the Fast Multipole
Algorithm (FMA) and ``classic'' tree--code to evaluate gravitational forces in
particle systems. 
We have optimized the Greengard's original
version of FMA allowing for a more efficient criterion of
{\it well--separation} between boxes, to improve the
{\it adaptivity} of the method (which is very important in
highly inhomogeneous situations) and to permit the {\it smoothing} of
gravitational interactions.

The results of our tests indicate that the tree--code is
almost three times faster than FMA for both a homogeneous and a clumped
distribution, at least in the interval of $N$
($N\leq 10^5$) here investigated and at the same level of accuracy
(error $\sim 10^{-3}$).
This order of accuracy is generally considered as the best compromise
between CPU--time consumption and precision for astrophysical simulation.
Moreover, the claimed linear dependence on $N$ of the CPU--time of FMA is not
confirmed and we give a ``theoretical'' explanation for that.
\end{abstract}



%
\section{Introduction}
The availability of fast computers is allowing a huge development of simulations
of large $N$--body systems in Astrophysics, as well as in other fields of
physics where the behaviour of large systems of particles have to be
investigated.

The heaviest computational part of a dynamical simulation of such systems
(composed by point masses and/or smoothed particles representing a gas) is the
evaluation of the {\it long--range} force, such as the gravitational one,
acting on every particle and
due to all the other particles of the system. 
Astrophysically realistic
simulations require very large $N$ (greater than $10^5$), making the direct
$\sim N^2$ pair gravitational interaction evaluations too slow to perform. 
To overcome this problem various approximated techniques to 
compute gravitational interactions have been proposed.\\ Among them,
the tree--code algorithm proposed by Barnes \& Hut (hereafter B.H.,
see [\cite{BH}]) is now widely used in
Astrophysics because it does not require any spatial fixed grid (like, for
example, methods based on the solution of Poisson's equation). This 
makes it apt to follow very inhomogeneous and variable in
time situations, typical of self--gravitating systems out of equilibrium.
In fact its intrinsic capability to give a rapid evaluation of
forces allows to dedicate more CPU--time to follow fast dynamical evolution,
contrarily to other higher accuracy methods that are more suitable in other
physical situations, e.g. molecular dynamics for polar fluids, where Coulomb
term is present.

While for the tree--code the CPU--time requirement scales as $N\log_8N$, in the
recently proposed Fast Multipole Algorithm (hereafter FMA, see
[\cite{greengard}])
this time is claimed to scale as $N$, at least in quasi--homogeneous
 2--D particle distributions. Were this linear behaviour confirmed
in 3-D highly non--uniform cases, FMA would really be appealing
for use in astrophysical simulations.\\ In this paper, we compare CPU times
of our own implementations of adaptive 3-D FMA and tree--code
to evaluate gravitational forces among $N$ particles in two rather different
(uniform and clumped) spatial configurations.

Detailed descriptions of tree--code and FMA can be found in
[\cite {BH}], [\cite {H}] and [\cite{greengard}] respectively. For the purpose
of
this paper (the performance comparison of the two above mentioned methods in
astrophysically realistic situations) we built our own computer versions of
tree--code and FMA. Our tree--code was written following at most the
[\cite {HK}] prescription but for the short--range component of the interaction
force (see [\cite {tesi}]), while our FMA 
is slightly different from the original proposed by Greengard
[\cite{greengard}], to make it more efficient in non--uniform 
situations where adaptivity is important, and to include the {\it smoothing}
of interactions which is quite useful in astrophysical simulations, as we will
describe in Sect. \ref{descrfma}.

In Sect. 2 and 3 we briefly review the algorithms and give some details of 
our implementations, in particular for the FMA, while in Sect. 4 the
comparison of FMA and tree--code CPU--time performances is presented and
discussed.
\section{The tree--code}
We have implementated our own version of the 
``classic'' BH {\it tree--code} (see [\cite{BH}] or [\cite{H}])
which is well described in [\cite{tesi}] (see also [\cite{HK}]);
we will give here only a brief discussion of the improvements we have
introduced. 

In the tree--code the cubic volume of side $\ell$ that
encloses all the particles is subdivided in 8 cubic {\it boxes} (in 3-D) and
each of them is further subdivided in 8 {\it children boxes} and so on.
The subdivision goes on recursively until the smallest boxes (the so called
{\it terminal boxes}) have  only
one particle inside. Moreover the subdivision is local and {\it adaptive} in
the sense that it is locally as more refined as the density is higher.
The logical internal representation of this picture is a ``tree data
structure'', from that the denomination {\it tree--code}.

The gravitational force on a given particle in $\bf r$ is then computed
considering the contribution of the ``clusters'' of particles contained in
boxes that are sufficiently distant from the particle, that is in those boxes
which satisfy the {\it open--angle} criterion:
\begin{equation}
\frac{\ell_m}{d}<\theta \label{opa}
\end{equation}
with $d$ the distance between the particle and the center of mass of the
cluster, $\ell_m=\ell/2^m$ the box size at level $m$ of refinement and
$\theta>0$ a parameter a priori fixed.
This contribution is evaluated by means of a truncated multipole
expansion that permits to represent
the set of $n$ particles contained in the box with a unique ``entity'' 
identified by a relatively low number of attributes (the total mass, the center of
mass, the quadrupole moment..., that is the set of multipole coefficients),
that are stored, in a previous step, in the tree data structure.
In this way the evaluation of interactions is speeded--up compared
to a direct pair-to-pair evaluation.
Usually in tree--codes second order expansions are used, that is up
to the quadrupole term, this approximation being sufficient in
typical astrophysical simulations.
Those boxes that do not satisfy the condition (\ref{opa}), will be
``opened'' and their {\it children} boxes will be considered. This ``tree
descending'' continues until one reaches a terminal box whose contribution 
to the field will be calculated {\it directly}. 

The parameter $\theta$ and the order of truncation of the expansion, permit
to have some control of the error made in the evaluation of the
field in the generic particle position. With the second order of approximation
we used and with $\theta$ in the interval $[0.7,1]$ one obtaines a relative
error less then about 1\%, as we will see in the following.

In our version of the tree--code we considered a {\it gravitational smoothing}.
This means that each particle is represented by a $\beta$-spline (a polinomial
function, with a compact support, differentiable up to the second order)
that gives a
Newtonian potential outside the sphere centered at the position
of the particle and of radius $2\epsilon$, while inside that sphere
the field is {\it smoothed}; see for example [\cite{HK}]
and [\cite{tesi}]. This smoothing
avoids divergence in the accelerations during too close approaches between
particles whose trajectories would be not well integrated in time by the usual
numerical procedures (like, for example, the {\it leap--frog} scheme). In
the particular case of our {\it static} test, i.e.
without considering any time evolution, the smoothing would only be an
unnecessary complication; consequently in these tests we
fix $\epsilon=0$ for every particle (exactly Newtonian potential).

However because of the presence, in general, of a smoothing of the field we
have modified the
open--angle criteria in the following way: the box of size $\ell$ with center
of mass at $\bf R$ is ``sufficiently distant'' from the particle in $\bf r$,
i.e. its potential can be expandend in a multipole series, if
\begin{equation}
|{\bf R}-{\bf r}|>\max\{\ell/\theta,2\epsilon\}
\end{equation}
where $2\epsilon$ is the gravitational smoothing ``radius'' of the particle in
$\bf r$.
\section{The Fast Multipole Algorithm}
\label{descrfma}
In a way similar to the tree--code, the FMA is based on an approximation
of the gravitational field produced by a set of sufficiently distant particles
over a generic particle. It uses the same logical
``octal tree--structure'' of the hierarchical subdivision of the space
as the tree--code, with the only difference that
instead of stopping subdivision when boxes contain single
particles, in this case the recursive subdivision ends up at
boxes containing no more than a fixed number $s>1$ of particles. This to
reduce the CPU--time required for the calculation. We still
call {\it terminal boxes} the boxes
located at the ``leaves'' of the logical tree--structure.
Also the FMA uses the truncated {\it multipole expansion}, but in a different
and more complicated form. The advantage is that of a
more rigorouse control of the truncation error.

In general the spherical harmonics expansion works for particles interacting
via
a potential $\Phi({\bf r},{\bf r}')\propto 1/|{\bf r}-{\bf r}'|$, so, even if
we will always speak in terms of gravitational field the
treatment is the same in the case of electrostatic field. The only differences
are the
coupling constant and the attribute of each particle (the mass in the
gravitational case, the charge with its sign in the electrostatic case).

The FMA takes advantage of that if we know the coefficients of the
{\it multipole expansion} that gives the potential in a point $P$, we are
able to build a Taylor expansion ---the so called {\it local expansion}--- of the
potential in a neighbourhood of $P$
(see Appendix \ref{fmatheo}).
This expansion is used to evaluate the acceleration of a particle near $P$,
instead of re-evaluating the multipole expansions of the field due to all
collections of distant particles as it happens with the tree--code.

Furthermore, besides the interaction
particle--particle and particle--box, FMA considers also an interaction
box--box which is estimated by a truncated multipolar expansion, if and only
if the boxes are sufficiently distant one each other, that is if they satisfy
the so called {\it well--separation} criterion (which substitutes the
open--angle criterion of the tree--code). Boxes which are not
well--separated, will be ``opened'' and their children boxes considered.
In this way one can control the error introduced by the truncation of the
multipolar expansion, having this error, as we will see, a well defined
upper bound.

In fact we know that, if we have
a set of $k$ particles at $P_i=(\rho_i,\alpha_i,\beta_i)$ (we will use
spherical coordinates) having masses $m_i$, which is enclosed in a sphere $A$
with 
center in the origin and radius $a$, and another set of $l$ particles at $Q_i=
(r_i,\theta_i,\phi_i)$ enclosed in a sphere $B$ centered
in $Q_0$ with radius $b$, then the approximated potential generated by the set
of particles in $A$ at the position of the $i$-th particle in $B$, is given
by a multipole expansion truncated at the order $p$:
\begin{equation}
\tilde{\Phi}_p(Q_i)\equiv -G\sum_{n=0}^p\sum_{m=-n}^n \frac{M^m_n}{r^{n+1}_i}
Y^m_n(\theta_i,\phi_i).
\label{mul1}
\end{equation}
The functions $Y^m_n(\theta,\phi)$ are the spherical harmonics and
\begin{equation}
M^m_n\equiv\sum_{i=1}^k m_i\rho_i^nY^{-m}_n(\alpha_i,\beta_i). \label{mul0}
\end{equation}
Now one can show (see [\cite{greengard}]) that, for any $p\geq 1$,
\begin{equation}
|\Phi(Q_i)-\tilde{\Phi}_p(Q_i)|\leq \frac{{\cal A}}{r_i-a}
\left(\frac{a}{r_i}\right)^{p+1}, \label{err1} 
\end{equation}
where $\Phi$ is the exact potential and ${\cal A}\equiv G\sum_{i=1}^k m_i$. 
This result gives the possibility to limit the error introduced by
approximating $\Phi$ with the expression (\ref{mul1}), if
the two sets of particles are sufficiently distant.

Let $d$ be the distance between the centers of the two spheres and let us define
$\sigma\equiv d-a-b$ the {\it separation} between
them. 
Obviously the set of particles in $B$ is such that
$r_i\geq a+\sigma$
for any $i=1,...,l$. 
Then from (\ref{err1}) we find:  
\begin{equation}
\Delta\Phi\equiv\max_{i=1,...,l} \left\{ |\Phi(Q_i)-\tilde{\Phi}_p(Q_i)|\right\}\leq
\frac{{\cal A}}{\sigma}\left(\frac{a}{\sigma+a}\right)^{p+1}.
\label{err2}
\end{equation}
The maximum error of the truncated multipole
expansion depends on two quantities: the
order of the expansion $p$ and the separation $\sigma$.

In his original algorithm, Greengard introduced the concept of
{\it well--separation} between
two boxes of the {\it same level} of refinement, i.e. with the same size
$\ell$, to control the truncation error.
Suppose to have 
two sets of particles in two boxes
circumscribed by two spheres $A$ and $B$, with equal radii $a=b=\ell\sqrt{3}/2$:
these two boxes are {\it well--separated} if, and only if, their separation
is such that $\sigma>a$, i.e. $d>3a$. In this way the upper bound on the error
will be, from eq. (\ref{err2}), 
\begin{equation}
\Delta\Phi\leq \frac{{\cal A}}{a}\left(\frac{1}{2}\right)^{p+1}.\label{errgre}
\end{equation}
To implement this criterion, Greengard imposed simply that:
two boxes of the same level are {\it well--separated} if, and only if, 
there are at least {\it two layers of boxes} of that level between them, in such
a way to
satisfy the inequality $d>3a$ and finally the (\ref{errgre}). 

Note that Greengard's well--separation criterion refers exclusively to boxes at
the same level of refinement, limiting the efficiency of the algorithm
especially in the case of non--uniform distributions of particles. 

We have modified this criterion to make it more efficient (in the {\it adaptive}
form of the algorithm) 
to face astrophysical typical distributions very far from uniformity.\\
Let us briefly explain our modification.\\
First of all, every box is associated to a sphere that is not merely the sphere
circumscribing the box but the {\it smallest} one containing all its particle.
So,
for terminal
boxes, is the smallest sphere concentric to the box and that contains
all its particles and
for a non--terminal box of size $\ell$, the sphere is also concentric to the
box but the radius $r$ is calculated recursively, knowing the radii $r_i$ 
$(i=1,...,8)$ of the spheres associated to each of its {\it children unempty
boxes}, via the formula
\begin{equation}
r=\max_{i=1,...,8}\{r_i\}+\ell\sqrt{3}/2. \label{radius}
\end{equation}
This sphere contains all the spheres associated to the children boxes\footnote
{This is to preserve the same error bound as in eq. (\ref{err3}), for the
truncated expansions obtained translating and composing
the local and the multipole expansions as discussed in Appendix \ref{fmatheo}}
. 
The value of the radius of these spheres is stored in the ``tree'' data structure together
with the other data pertinent to the box and it is much more
representative of the ``size'' of the set of particles in the box, than
the mere box size $\ell$, for controlling the truncation error.

\mettips{fig2}{}{}{}{}{Potential field produced by $k$ particles on $l$ other
ones.}{Example of two {\it well--separated} boxes, $\hat A$ and $\hat B$
(see text). $A$ and $B$ are the associated spheres enclosing the two set of
particles.}{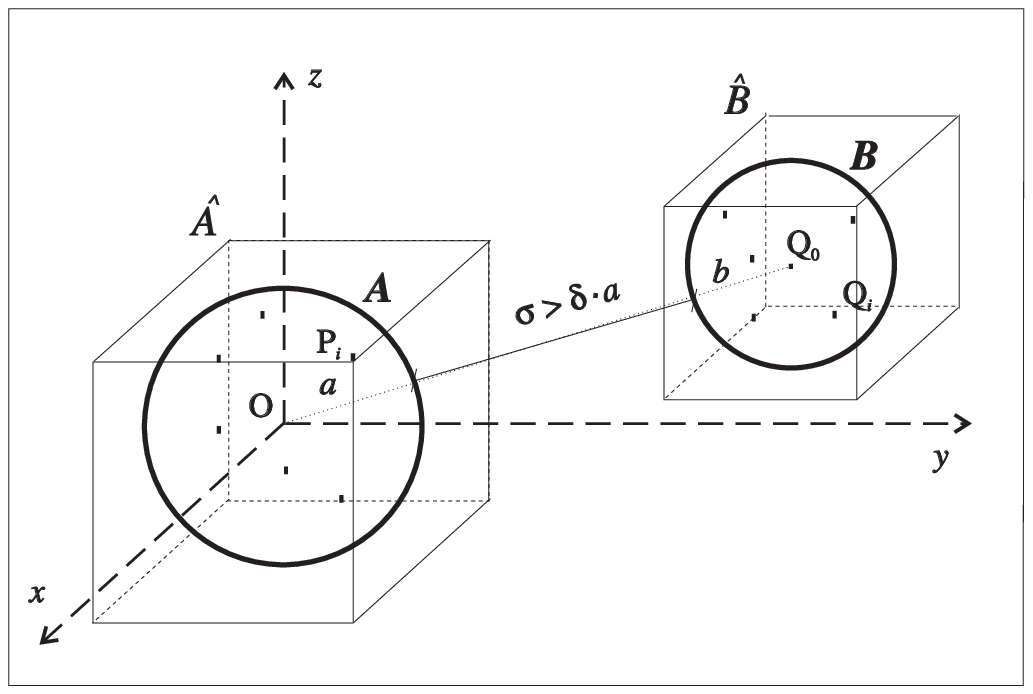}
Now consider the box $\hat A$ centered in the origin $O$ and containing the 
collection of $k$ particles belonging to the sphere $A$ seen before.
Let this sphere $A$ be its associated sphere with radius $a$ (see Fig.
\ref{fig2}).
%
Let the other set of particles at $Q_i, i=1,..,l$, be enclosed 
in the box $\hat B$ whose associated sphere be the sphere $B$ 
centered in $Q_0$ and with radius $b$. The radii $a,b$ have been evaluated
by the (\ref{radius}).
Let us suppose that the separation of the two spheres is such that 
$\sigma>\delta\cdot a$, with 
$\delta >0$ a fixed parameter (see again Fig.\ref{fig2}).
Obviously the set of particles in $B$ is such that, from eq. (\ref{err2}),
\begin{equation}
\Delta\Phi\leq
\frac{{\cal A}}{\delta\cdot a}\left(\frac{1}{\delta+1}\right)^{p+1}
\label{err3}
\end{equation}
and we can note that the parameter $\delta$  
works in a way similar to $\theta$ in the tree--code.

Hence, our criterion of {\it well--separation} is the following: once we fixed
the parameter $\delta$, we define two boxes
as {\it well--separated}, if the distance between their centers $d$ is
such that
\begin{equation}
d>a+b+\delta\cdot a, \label{welsep}
\end{equation}
that is if the separation between their
associated spheres is $\sigma>\delta\cdot a$. In this way the evaluation of the
interaction between the 
sets of particles contained into such boxes will be affected by a truncation
error which is bounded by the (\ref{err3}).

The implementation of this
criterion is very easy because it consists just in verifying the (\ref{welsep})
where the radii of the spheres associated to the boxes have been already
calculated and stored, together with 
all the other data of each box, in the phase of the algorithm in which the
tree--structure is built.  

Our criterion of {\it well--separation}
is more efficient than the Greengard's original one, having the following
features:
\begin{itemize}
\item it depends upon the internal distribution of particles in the box (via
the radius of the associated sphere);
\item it can involve boxes of different level of refinement, having different
sizes, so to improve efficiency in non--uniform situations and to make
unnecessary those complicated tricks
conceived to shorten the list of the well--separated boxes like, for
example, the
mechanism of ``parental conversion'', (see [\cite{board}]);
\item it can be tuned in such a way to obtain the desired accuracy, via the
parameter $\delta$;
\item it can be easily modified in order to allow a {\it gravitational
smoothing} to be applied to the particles, as we will see below.
\end{itemize}

Note also, see the (\ref{errgre}), that to control the truncation error,
Greengard varied $p$
according to the desired accuracy.
In our version we have, instead, fixed $p=2$ and achieved
varied the parameter $\delta$ in order to obtain the same accuracy that is
usually obtained with the tree--code in astrophysical simulations.
One has to care with keeping reasonably low the execution CPU--time (be
small compared with human time scale!) and this
is obtained at a price of a certain loss in accuracy in the evaluation of
interactions.
In fact the main characteristics of astrophysical simulations of gravitating
systems are:
\begin{enumerate}
\item they require a great number of particles
(usually more than $10^4$)
for a rather large duration of the simulation\footnote{several {\it dynamical
times}: that is many times the typical time scale of the entire system, such as 
the sound crossing
time for a collisional system or the core--crossing time for a collisionless
one}
and, in particular,
\item due to the intrinsic instability
they offer a very wide distribution of time scales.
\end{enumerate}

So while simulating polar fluids in molecular dynamics, one has to face
``microscopic'' time scales more narrowly distributed (just because molecules
tend to repulse each other due to the presence of short--range interactions,
like the Lennard--Jones potential) and one can work with expansions
truncated up to eighth order or more (although the CPU--time grows as $p^4$,
see [\cite{greengard}]), in astrophysical simulations one prefers to limit the
precision at a lower but reasonable level and, on the other hand, to be able
to process systems which are highly dynamical. Therefore one usually works
with expansions truncated up to the second order (in some cases even the
first order) that, in the tree--code, corresponds to consider up to the
quadrupole moment.

For the mass density of the single particle we used the same {\it $\beta$--spline}
profile as we did in our tree--code. 
In this way the potential is exactly Newtonian outside the sphere
of radius\footnote{In principle each particle is allowed to have its own 
smoothing length. This is useful when one has to simulate
gravitational interactions between both {\it collisionless} and {\it
collisional} particles (i.e. fluid elements).} $2\epsilon_i$ centered in
${\bf r}_i$, so it can be expanded in
multipole series {\it only in this region}. Hence the interaction with a
particle inside the sphere of radius $2\epsilon_i$ must be necessarily
evaluated by means of a {\it direct} summation.
This requirement, which would be very difficult to incorporate in Greengard's
criterium, has been considered via a little arrangement of
our {\it well--separation} criterium (\ref{welsep}), that is:\\ two boxes
($\hat A$ and $\hat B$) are {\it well--separated} if, and only if, their
spheres ($A$, with radius $a$ and $B$ with radius $b$ respectively) are such
that:
\begin{equation}
d\geq a+b+\max\{\delta\cdot a,2\epsilon_{\hat A}\}
\end{equation}
where $d$ is the distance between the centers of the two spheres.

The smoothing length
$\epsilon_{\hat A}$ used for the box $\hat A$ is another quantity stored in
the tree data structure and it is given by this simple recursive scheme:
if $\hat A$ is terminal, then $\epsilon_{\hat A}\equiv\max_i\{\epsilon_i\}$,
where $\epsilon_i$ is the smoothing lenght of the particle $i$ contained in
$\hat A$, otherwise
$\epsilon_{\hat A}\equiv\max_{\hat C}\{\epsilon_{\hat C}\}$,
with the maximum taken over all the unempty children boxes $\hat C$ of
$\hat A$.
Anyway in the following comparison tests, we let $\epsilon=0$ (as for the
tree--code) because we are not interested in the dynamical evolution of the
system.

For a more detailed
description of our own implementaton of FMA see Appendixes \ref{fmatheo} and
\ref{formal}. 
\section{Codes performance comparison}
To compare the CPU--time spent by the two algorithms described in
the previous Sections, we ran the codes on a IBM R6000 workstation with
two different distributions of particles.
In the first case $N$ particles have been distributed
a {\it uniformly} at random in a sphere of unitary radius.
In the second case a set of $N$ particles has been distributed, with a
Monte--Carlo method, 
in a unitary sphere in such a way to discretize the density
profile:
\begin{equation}
\rho(r)=\frac{\rho_0}{\left[1+\left(\frac{r}{r_c}\right)^2\right]^{5/2}},
\end{equation}
with $r_c=0.2$
(obviously $\rho=0$ for $r>1$). This latter is known as Schuster's ([\cite{schuster}]) profile; it
corresponds to a polytropic sphere (of index 5) at equilibrium (see
[\cite{Binney}])
and represents a good approximation to the density distribution of various
stellar systems. In both the uniform and clumped case all the particles are
assumed to have the same mass.

The order of accuracy chosen was the same for both the tree--code and the FMA. 
For accuracy we mean how close, in modulus, the evaluated forces are to those calculated
``exactly'' by a direct, {\it Particle--Particle} (PP) method, which is
affected only by the numerical error of the computer (due
to the finite number of digits). Consequently we define as {\it relative
error} of the calculation $\varepsilon$:
\begin{equation}
\varepsilon\equiv\frac{1}{N}\sum_i^N\frac{|a_i -a^{PP}_i|}{a^{PP}_i},
\end{equation}
where $a_i$ is the modulus of the acceleration of the $i$-th particle
estimated by each of the two algorithms and $a^{PP}_i$ that computed by the PP
method.
The error on the {\it direction} of the forces is much lower then the error
on the modulus we have defined above, and, as it is usual in N--body numerical
method, it is not considered at all in performance tests being negligible.
\epsfxsize 5 truein
\mettips{fig.err}{}{}{}{}
{relative error}{Averaged relative error vs. number of particles for both
algorithms.}{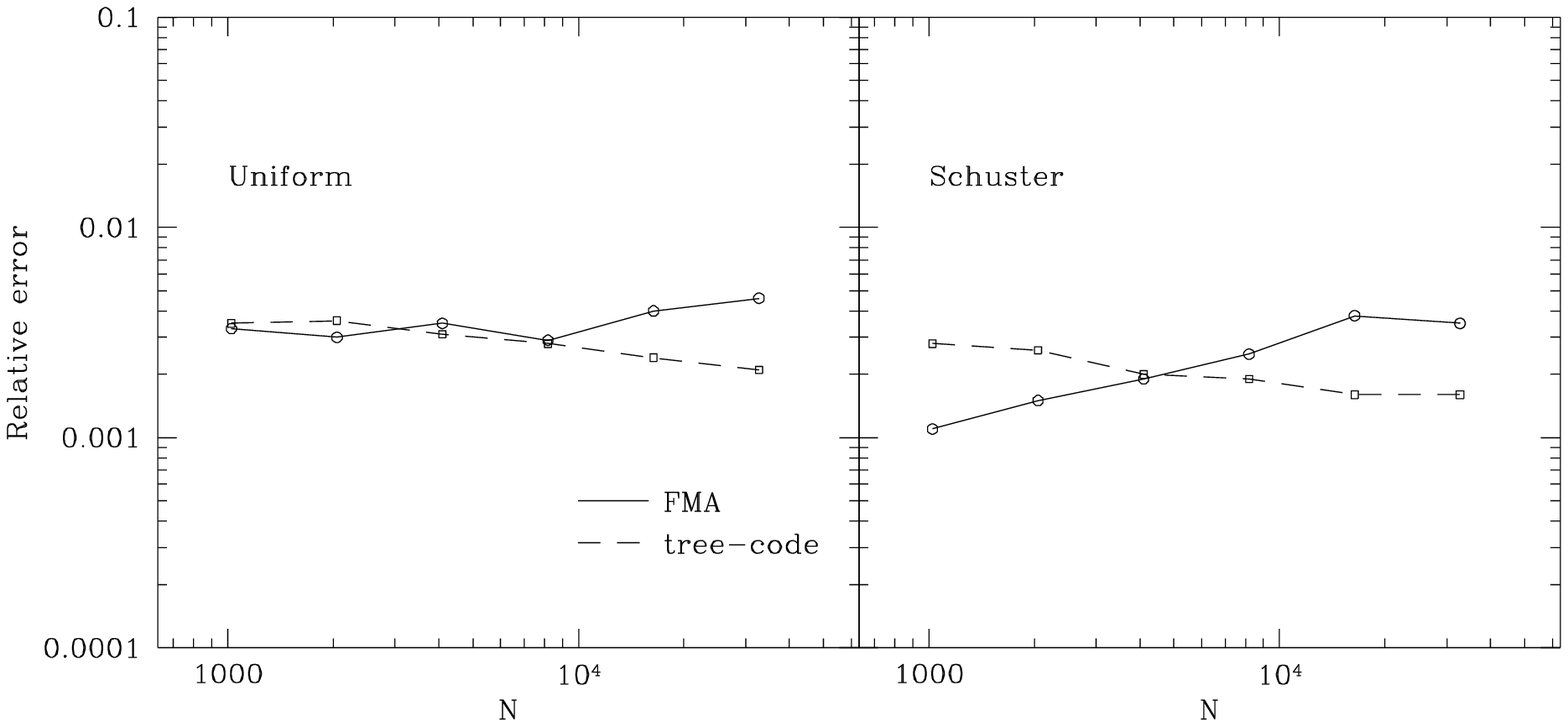}
\epsfxsize 5 truein
\mettips{fig.time}{}{}{}{}
{CPU--time vs. number of particles}{CPU--time (on an IBM R6000 machine) vs.
number of particles for both
the algorithms and for the direct Particle--Particle method.}
{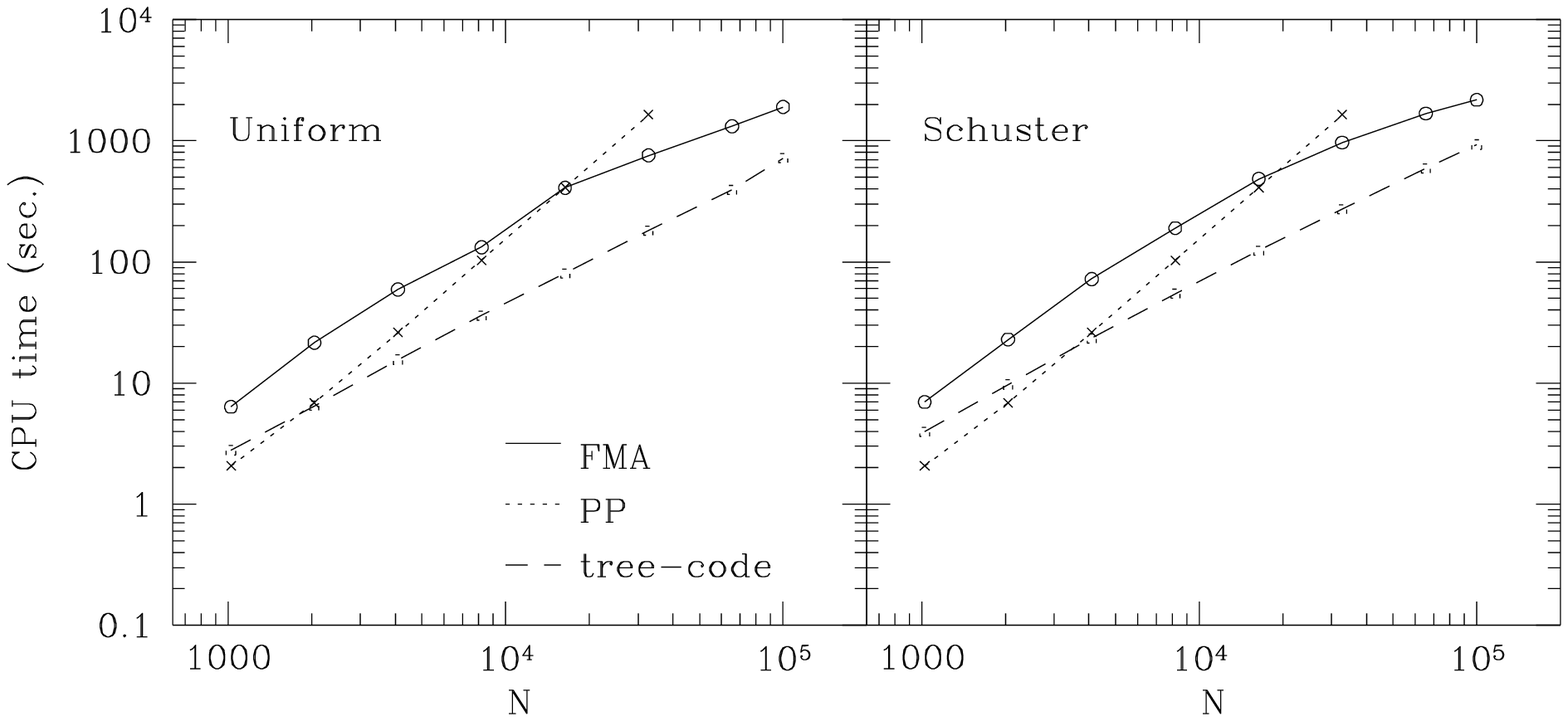}

Figure \ref{fig.err} gives the relative error of
the tree--code and of the FMA in the uniform and ``clumped'' case.
The error is almost the same for both the
algorithms.
An averaged (on all the particles) relative error
less than 1\% (this is the order of magnitude of the error
generally admitted in astrophysical simulations), is obtained fixing
\begin{eqnarray}
\theta&=&0.7 \mbox{ for the tree--code,}\\
\delta&=&2.5 \mbox{ for the FMA},
\end{eqnarray}
and considering, as we have said, up to the quadrupole term in the tree--code
and to the same order term in the multipole expansion ($p=2$) of the
FMA. Moreover we chose $s=10$ in the FMA as the best compromise between accuracy
and computational speed, as we checked.

The CPU--time spent to calculate the accelerations vs. the number $N$ of
particles is shown in Fig.\ref{fig.time}. The CPU--time for the {\it
Particle--Particle} method is also shown as reference.
Both the algorithms are slower to compute
forces in the non--uniform model than in the uniform one, and for the tree--code
this is more evident.
This is clearly due to the more complicated and non-uniform spatial
subdivision in boxes that affects mostly the tree--code due to the {\it
finer} and {\it deeper} subdivision of the space it uses.
Anyway 
the tree--code shows to be faster than the FMA for both the distributions and
for $N$ varying in the range we tested.  
As expected, the behaviour of the CPU--time vs. $N$ for the tree--code is well
fitted by the logaritmic law
\begin{equation}
t_{cpu}=\alpha N\log N+\beta, \label{tlog}
\end{equation}
where $\alpha$ and $\beta$ are given in Table \ref{tavola}.
In our opinion, this law must be followed by the FMA too, as we
will explain in Section \ref{scaling}.

However, let us observe the CPU--time for the uniform case: it is not easy to
distinguish at a first sight a logarithmic
behaviour from a linear one, furthermore we can presume, as we can observe in
Figure \ref{fig.time}, that the FMA must show a more complicated behaviour due to the
presence of the parameter $s$ (the maximum number of particles leaved in
terminal boxes) and to that $s$ was kept {\it fixed}.
Hence for a certain range of $N$
this $s$ could have been chosen as optimum, but could have not been so for
other values of $N$
(in those ranges in which the CPU--time shows to grow excessively with a
slope larger than that of the tree--code and comparable to that of the PP
method).
Blelloch and Narlikar [\cite{BN}] have obtained similar
``undulations'' in the behaviour of the CPU--time of their version of FMA,
while the behaviour of tree--code was much ``cleaner'', as in our tests.

Coming back to the accuracy, note that for $N>8192$, the rel. error
of the FMA exceeds
that of the tree--code (see Fig.\ref{fig.err}). In the same range of $N$ the
CPU--time of the
FMA grows less rapidly than that of the tree--code. This
follows the obvious gross rule by which the faster the calculation 
the lower the precision in the results.
So for $N>8192$ the value $s=10$ fixed is likely too
small and consequently the percentage of {\it direct} (PP)
calculation is decreased, so that FMA loses precision. On the contrary, for
$N<4096$ we
observe, especially in the Schuster case, that
$\varepsilon_{FMA}<\varepsilon_{tc}$; in the
same region the CPU--time for FMA grows more rapidly than for the tree--code,
so in this case $s$ seems to be too high.

\begin{table}
\begin{center}
\begin{tabular}{cc}
Uniform & Schuster \\
\begin{tabular}{|c|c|c|}
\hline
Algorithm & $\alpha$ & $\beta$ \\[0.8ex]
\hline
tree--code &$1.1\cdot 10^{-3}$&$-1.1$ \\
FMA &$4\cdot 10^{-3}$&$-7.4$ \\
\hline
\end{tabular}
&
\begin{tabular}{|c|c|c|}
\hline
Algorithm & $\alpha$ & $\beta$ \\[0.8ex]
\hline
tree--code &$1.6\cdot 10^{-3}$&$-1.7$ \\
FMA &$4.9\cdot 10^{-3}$&$-10$ \\
\hline
\end{tabular}\\
\end{tabular}
\end{center}
\caption{Values obtained for the parameter, fitting the
behaviour of the CPU--time for both algorithms with the law:
$t_{cpu}=\alpha N\log_8 N+\beta$.\label{tavola}}
\end{table}
It is interesting to note that the relative error of the tree--code shows a
decrease at
increasing the number of particles. This is probably due to the large fraction
of {\it direct, particle--particle} calculations, because of the
growing density of particles (the volume occupied by the system
remains the same).
This explains also the lower error in the clumped case than in the
uniform one at a given $N$, because in the Schuster's profile the central
density of particles is clearly higher than in the uniform case.

Finally we can see that our FMA becomes faster than the PP method
for $N>18,000$ in the Schuster's model, and
for $N>16,000$ in the uniform case.\\ To make a comparison with codes of other
authors, let us consider the FMA implemented in 3-D by Schmidt and Lee
[\cite{SL}] (although this code {\it is not adaptive}) 
following the original
Greengard's algorithm and, in particular, his well--separation criterion.
Their FMA is {\it vectorized} and it runs on a CRAY Y-MP, but anyway we
do not make a direct comparison,
but rather compare our ``CPU--time ratio'' (that is the ratio between the
CPU--time consumed by FMA and that consumed by the direct PP method),
with the same ratio as obtained by the codes of Schmidt and Lee,
{\it at the same order of magnitude of the error on the forces}.

So we can note (see [\cite {SL}]) that for $N=10^4$ uniformly distributed
particles, a truncation of eighth order ($p=8$) and five levels of refinement
({\it a priori} chosen),
they obtain a CPU--time of 418 sec. for their FMA, against 11 sec. spent by
the {\it direct} PP method, with a relative error on the forces of about
$1.4\cdot 10^{-3}$.
Thus they obtain a ratio $t_{cpu}^{(FMA)}/t_{cpu}^{(PP)}\sim 40$, while with
our codes we see that for the uniform distribution with $N=10^4$, we have
an error on
the forces that is $\sim 3\cdot 10^{-3}$ (see Fig.\ref{fig.err}) with a
CPU--time ratio $t_{cpu}^{(FMA)}/t_{cpu}^{(PP)}\sim 1$, i.e. about the same
CPU--time spent by the PP method (see Fig.\ref{fig.time}).

This seems a very good result.
\subsection{Scaling of CPU--times versus $N$}
\label{scaling}
Greengard and other authors ([\cite{greengard}],
[\cite{board}],[\cite{salmon}]) assert that FMA would exhibit a {\it linear}
scaling of CPU--time vs. $N$. 
We tried to fit $t_{cpu}^{(FMA)}$ as function of $N$, with various laws,
the linear included. The result is that a {\it logarithmic} behaviour
like that of eq. (\ref{tlog}), gives the best fit, as for the tree--code.

The values obtained for $\alpha$ and $\beta$ in Table \ref{tavola}
show that $\alpha_{FMA}/\alpha_{tc}\sim 3$: the tree--code is roughly
{\it three times faster} than the FMA.
Note how this difference of performances
reduces slightly passing to the Schuster clumped profile; this because, as we
have said, our FMA adaptive code is less sensitive to the degree of uniformity
of the distribution of the particles than the tree--code which uses a
finer subdivision of the space in boxes (as it corresponds to $s=1$).

The higher speed of the tree--code  respect to the FMA
is easily understood, at least in 3-D, since while in
theory the FMA
is more efficient and less ``redundant'' in managing informations (remember
the use of Taylor expansion of the potential on near bodies), in
practice this ``potential'' greater efficiency pays the price of a certain quantity
of computational ``complications''. This carries the method to a negative total
balance in terms of speed respect to the competing tree--code.

How can we interpret the CPU--time scaling?\\ The logarithmic behaviour of
the tree--code is explained by a simple estimate of the
number of operations needed by the various steps of the algorithm
(see e.g. [\cite{BH}],[\cite{H}] and [\cite{tesi}]). It is roughly given 
by the product between the number of particles $N$ by the number of ``bodies''
(boxes or particles), about $\log_8 N$,
which contribute to the force on each particle.
Hence $t_{cpu}^{(tr)}\sim N\log_8 N$.

The logarithmic behaviour of the FMA can be similarly understood when
one reconsiders carefully the cost of each step of the algorithm.\\
It can be estimated (see [\cite{greengard}])  that 
\begin{equation}
t_{cpu}^{(FMA)}\sim aN+bN_{ter}+cN_{ne},\label{tcpufma}
\end{equation}
that is the CPU--time is a {\it linear function} of $N$, 
$N_{ter}$ (the number of all {\it terminal} boxes) and $N_{ne}$ (the 
number of all non--empty boxes). 
Greengard [\cite{greengard}] in his final considerations on the scaling of the
FMA in the adaptive 2-D version (the 3-D case is similar),  
estimates the number of this
types of boxes (see lemmas 2.6.4
and 2.6.5 in [\cite{greengard}]) to be
\begin{eqnarray}
N_{ter}&\sim& 4\cdot L\cdot\frac{N}{s} \label{nter}\\
N_{ne}&\sim& 5\cdot L\cdot\frac{N}{s} \label{nne}
\end{eqnarray}
where $L$ is the total {\it number of subdivisions} needed to reach
terminal boxes.
This $L$ is identified by Greengard with $L\sim\log_2 (1/\Delta)$,
where $\Delta$, {\it a priori} fixed, is the {\it spatial resolution} that
one wants to reach in the simulation.
Being $\Delta$ fixed, $L$ would be {\it independent} of $N$, thus
$N_{ter}$ and $N_{ne}$, in eq. (\ref{tcpufma}), are quantities
linear in $N$. Then the FMA CPU--time estimated by the
eq. (\ref{tcpufma}) results linear in $N$ too.

The crucial point is that a constant $\Delta$ (and $L$) allows to manipulate
only those distribution of particles such that $\Delta< r_{min}$, with
$r_{min}$ the minimum distance between a pair of particles (see observation
2.5.1 in [\cite{greengard}]). 

Obviously, in 3-D, $r_{min}\sim N^{-1/3}$, so that $\Delta<r_{min}$ implies  
\begin{equation}
L\undersim{>}\frac{1}{2}\log_2 N=\log_8 N \label{logN}
\end{equation}
Substituting this inequality in the expressions (\ref{nter}) and (\ref{nne}) for
$N_{ne}$ and $N_{ter}$, one obtains from eq. (\ref{tcpufma}) that
$t_{cpu}^{(FMA)}\undersim{>} N\log_8 N$, that is the same ``natural'' scaling of
the tree--code.
\section{Conclusions}
Realistic astrophysical simulations are characterized by the large amount of
computations required by the evaluation of gravitational forces. Many codes
which give performant approximation of the force field have been
proposed in the literature.
In this paper we compare two of these codes: one (the BH tree--code) has been
largely used in astrophysics for ten years, the other (the Greengard's FMA
[\cite {greengard}]) has 
given promising results in the field of molecular dynamics.\\
We have so implemented our own optimized {\it serial} versions of both the
tree--code and {\it adaptive}, 3-D, FMA.

The results of our comparison tests indicate the tree--code as faster than
FMA over all the interval of total number of particles ($N\leq 10^5$) allowed
by the central memory capacity of a ``typical'' workstation.
This maximum value of $N$, which could appear low respect to modern parallel
simulations
(up to $N\sim 10^7$), is anyway meaningful because even fully parallel codes
are limited by an {\it individual} processor charge of that order of
magnitude. The problems in the parallelization of the two codes are comparable
due to their similar structure, thus the higher speed of the tree--code, here 
verified in a serial context, should be confirmed in the parallel
implementation and it seems
a valid reason to concentrate efforts for the most efficient parallelization
of the tree--code.

At the end of this paper we have discussed the dependence of
the FMA CPU--time on $N$ and given explanation of why its behaviour is
similar to that of the classic tree--code.
\appendix
\section{Manipulation of Multipole Expansions in the FMA}
\label{fmatheo}
Here we briefly describe the three theorems, due to Greengard [\cite{greengard}],
that
permits the manipulation, in 3-D, of the various series expansions used in the
algorithm, and that are useful for the deeper and formal description of our own
implementation that follows in the next Appendix \ref{formal}.

We have said that one can ``transform'' the multipole expansion (\ref{mul1})
into a {\it local expansion} useful to evaluate the field about a given
point $P$.
More precisely, given the set of $k$ particles at $P_i$ in the sphere 
$A$ (associated to the box $\hat A$, see the Fig.
\ref{fig2}) 
which produce the gravitational potential $\Phi(Q_i)$ over the set
of particles in the sphere $B$ (box $\hat B$) with center $Q_0$, then
one can show that in the vicinity of $Q_0$ the approximated potential
\begin{equation}
\Psi_p(Q_i)=-G\sum_{j=0}^p\sum_{k=-j}^j L_j^k Y_j^k(\theta'_i,\phi'_i)(r'_i)^j,
\label{tay1}
\end{equation}
where $Q_i-Q_0=(r'_i,\theta'_i,\phi'_i)$, differs from the exact $\Phi(Q_i)$ of 
an amount bounded by the same expression that appears in the r.h.s. of the
eq. (\ref{err3}).  
In this truncated {\it local
expansion} the coefficients are given by:
\begin{equation}
L_j^k=\sum_{n=0}^p\sum_{m=-n}^n S_{j,k,n,m}^{(1)}M_n^mY_{j+n}^{m-k}(\alpha_i,
\beta_i)(\rho_i)^{-j-n-1}, \label{multay}
\end{equation}
$M_n^m$ being the {\it same} that appears in the expression (\ref{mul1}) and
$S^{(1)}$ a matrix of coefficients (see Appendix \ref{coeff}). 

Another theorem allows us to calculate $M_n^m$ in a {\it recursive} manner.
Let us consider the {\it partition} of the set of $k$ particles in the
box $\hat A$, in the $q\le 8$ sub--sets each of them enclosed in the spheres
associated to the children boxes of $\hat A$ and with centers in
$(\rho^{(1)},\alpha^{(1)},\beta^{(1)}),...,
(\rho^{(q)},\alpha^{(q)},\beta^{(q)})$.
Let ${M_n^m}^{(1)},{M_n^m}^{(2)},...,{M_n^m}^{(q)}$ be the multipole
coefficients calculated for each of the sub--sets of particles.
If all these spheres are enclosed in the
sphere $A$ (this is automatically satisfied because of the eq.
(\ref{radius})),
the coefficients $M_j^k$ given by:
\begin{equation}
M_j^k=\sum_{n=0}^j\sum_{m=-n}^n S_{j,k,n,m}^{(2)}{M_{j-n}^{k-m}}^{(q)}Y_n^{-m}
(\alpha^{(q)},\beta^{(q)})(\rho^{(q)})^n,\label{mulcom}
\end{equation}
(see the Appendix \ref{coeff} for the matrix $S^{(2)}$) give a 
potential
\begin{equation}
\tilde{\Phi}_p(P)=-G\sum_{n=0}^p\sum_{m=-n}^n \frac{M^m_n}{r^{n+1}}
Y^m_n(\theta,\phi)
\end{equation}
(where $P=(r,\theta,\phi)$ is a generic point {\it outside} the sphere
$A$), which {\it well approximates} the exact potential
$\Phi(P)$ generated by {\it all} the $k$ particles in the sphere $A$. In fact
if $P$ is the position of a particle in the {\it well--separated} box 
$\hat B$, then one can show that:
\begin{equation}
|\tilde{\Phi}(P)-\Phi(P)|\leq \left(\frac{G\sum_i^k m_i}{r-a}\right)
\left(\frac{a}{r}\right)^{p+1}<
\frac{G\sum_i^k m_i}{\delta\cdot a}\left(\frac{1}{\delta+1}\right)^{p+1},
\label{err4}
\end{equation}
the truncation error has the same upper bound given by (\ref{err3}).

So we can compute $M_j^k$ associated to the box $\hat A$ containing the total
set of particles knowing only those pertinent to the $q$ sub--sets in each
children box.
In the tree--code the same happens, but there the analogous theorem---the  
``quadrupole composition theorem''---has been developed {\it specifically} for
the quadrupole moment by Goldstein [\cite{Goldstein}] (those for the monopole
and the dipole are obvious). In the FMA this theorem works
for coefficients of any order and it has an upper error bound.

Thus, once $M_n^m$ have been calculated for all {\it terminal} boxes
using the definition (\ref{mul0}), by means of (\ref{mulcom}) we can compute
recursively the coefficients of {\it parent} boxes ascending
the tree--structure. This coefficients
will be transformed, when needed, in the local expansion coefficients (as we
will see in more details in Appendix \ref{formal})
necessary to calculate forces by (\ref{tay1}).
In this way, we will be sure that the error made in
approximating the ``true'' potential with the various expansions,
will always be bounded by the (\ref{err4}).

The last theorem concerns with the translation and composition of the {\it
local expansion} coefficients (briefly {\it Taylor coefficients}). In this
case the rules of composition works, in a certain sense, inversely.
That is, given the coefficients $L_j^k$ relative to the set of $k$ particles
in the sphere $A$, such that the potential $\Psi_p(P)$
($P=(r,\theta,\phi)$ is a point {\it inside} $A$) given by the truncated
local expansion about the origin $O$:
\begin{equation}
\Psi_p(P)\equiv-G\sum_{j=0}^p\sum_{k=-j}^j L_j^k Y_j^k(\theta,\phi)r^j,
\label{tay2}
\end{equation}
is such that
\begin{equation}
|\Psi_p(P)-\Phi(P)|<\frac{G\sum_i^k m_i}{\delta\cdot a}
\left(\frac{1}{\delta+1}\right)^{p+1},
\end{equation}
then at the same point but with another origin $Q\in A$, we have the equality
\begin{equation}
\Psi_p(P)=-G\sum_{j=0}^p\sum_{k=-j}^j {L^Q}_j^k Y_j^k(\theta',\phi')(r')^j
\label{tay3}
\end{equation}
where $P-Q=(r',\theta',\phi')$ and where the new translated coefficients are:
\begin{equation}
{L^Q}_j^k=\sum_{n=j}^p\sum_{m=-n}^nS^{(3)}_{j,k,n,m}L_n^mY_{n-j}^{m-k}
(\alpha,\beta)\rho^{n-j}\label{Taytrans}
\end{equation}
being $O-Q=(\rho,\alpha,\beta)$ (for the matrix $S^{(3)}$ see Appendix
\ref{coeff}). 

Thus given the $L_j^k$ for a box, we can compute the Taylor coefficients for
all the unempty {\it children boxes} using the above formula, with the new
origin $Q$ at the center of each children boxes. The process has to be
recursively iterated until we reach terminal boxes.
But how can we obtain the coefficients $L_j^k$ of a box $B$ ``the first 
time'' (not knowing those of its parent box)?
Obviously they will be calculated by means of the (\ref{multay}), transforming
the multipole coefficients of {\it sufficiently distant} boxes. That is of
boxes that are {\it well--separated} from $B$.
\section{Formal description of the FMA algorithm}
\label{formal}
Here we describe in deeper detail our version of the FMA algorithm
that
is sligthly different from the original Greengard's
algorithm in the adaptive implementation. The differences
regard mainly the way interactions between distant boxes and the set of
particles in a terminal box are calculated. Moreover, as we have said, we
have modified the Greengard's {\it well--separation} criterion to take
into account the presence of a {\it smoothing} of the interaction that in
astrophysical simulations, contrarily to molecular dynamics, is
unavoidable to include.

Let us first
introduce some useful definitions: in the following $s$ indicates
the maximum number
of particles in the {\it terminal boxes} and the ``calligraphic'' letters
refer to collections of boxes while simple capitals letters refer to single
box.
\begin{itemize}
	\item $l_{max}$ is the maximum level of refinement reached in the
space subdivision;
	\item $l_A$ indicates the level of box $A$, whereas the level of the
 {\it root box} $R$, that is the box containing all the particles, is $0$;
	\item ${\cal B}(l)$ is the set of all boxes at level $l$ of refinement;
	\item $M(A)$ is the {\it parent} box of box $A$;
	\item ${\cal C}(A)$ is the set of all {\it children boxes} of box $A$;
	\item ${\cal C}({\cal S})\equiv\cup_{A\in{\cal S}}{\cal C}(A)$ is the
set of all children boxes of each box in the set $\cal S$; 
	\item ${\cal X}(A)$ indicates the set of boxes, of level $l_A$ or
$l_A+1$, that are {\it NOT well separated} from box $A$. The set
contains all the {\it brothers} of box $A$, but not $A$ itself;
	\item $\cal T$ is the set of {\it terminal boxes}, that is such boxes
that have no children because they contain less than $s+1$ particles,
so they have not been subdivided;
	\item $n_A$, with $A\in {\cal T}$, is the number of particles inside
the terminal boxes $A$ (obviously $n_A \leq s$);
	\item $d_{AB}$ represents the distance between the geometrical centers
of boxes $A$ and $B$;
	\item $\epsilon_A$ is the length of the {\it gravitational smoothing}
relative to the box $A$;
	\item $r_A$ is the radius of the sphere that contains all the
particles in the box $A$ and that is concentric to it (see text).
\end{itemize}

The notation `do $n=a,b$' (with $b>a$ integers) means that all passages
included between this statement and the correspondent `end do', are repeated
$b-a+1$ times and every time the integer variable $n$ takes the values:
$a,a+1,...,b-1,b$, like in the Fortran, while the notation
`do $A\in {\cal S}$' means, in this
case, that every time the loop is executed the {\it box} $A$ represents one of
the various boxes in the set $\cal S$. 
So the statements between `do'
and the related `end do' are repeated Card$\{{\cal S}\}$ times and each time
with a different box $A\in {\cal S}$. For example,
if ${\cal S}=\{ A_1,A_2,A_3,...,A_n\}$, the box $A$ is $A_1$ the first time
the loop is executed, $A_2$, the second time and so on.
However, the order the boxes have in the set
$\cal S$ has no importance in the algorithm. On the contrary in
the first case of `do ... end do', the order in the values that $n$
takes everytime is important.
Another notation is `do while {\it condition}', meaning that it will be
executed the statements between this `do while ...' and the correspondent
`end do', {\it while} the logical condition keeps {\it true}.
{\small
{\parindent 0 pt

{\leftskip 0.5 true cm \rightskip 0.5 true cm
{\bf Calculate} recursively the multipole coefficients for all the boxes
of each level, starting from the terminal boxes. This procedure is the same
that in the tree--code,
but with the difference that in the FMA the multipole expansion is calculated
with the origin in the geometrical center of the boxes, so the first order
coefficient (the dipole moment) does not vanish. Another
complication is that in the FMA this coefficients are necessarily complex
quantities. For terminal boxes use the eq. (\ref{mul0}), while for the others
use eq. (\ref{mulcom}).

{\bf Calculate} the radius $r$ of the sphere that contains
all the particles in each box. If a box is {\it terminal} then
$r\equiv\max_i\{|{\bf r}_i-{\bf R}|\}$, where ${\bf r}_i$ is the position of
the particle $i$ in the box and $\bf R$ is its center.
If it is not a terminal box the radius is calculated by means of
(\ref{radius}).

{\bf let ${\cal X}(R)=\O$}

{\bf do $l=1,l_{max}$}

{\leftskip 1 true cm
  {\bf do $C \in {\cal B}(l)$}

  {\bf let ${\cal X}(C)=\O$}

  {\leftskip 1.5 true cm
    {\bf if $l>1$ then} translate, if they exist, the coefficients of the
    local expansion of the
    parent box $M(C)$ about the center of box $C$ using eq. (\ref{Taytrans}).

    {\bf if $C \not\in {\cal T}$ then} [the box $C$ isn't terminal]

    {\leftskip 2 true cm
      {\bf do $B \in {\cal X}(M(C))$}

      {\leftskip 2.5 true cm
	{\bf if $d_{BC} \geq \max\{2\epsilon_B,\delta\cdot r_B\} + r_B+r_C$
	       then}

	{\leftskip 3 true cm
	  {\bf convert} the multipole coefficients of box $B$ 
	  to Taylor coefficients about the center of box $C$ with eq.
	  (\ref{multay}), because it is {\it well separated} from $C$ and
	  the sphere where the field generated by the masses in $B$ is {\it
	  smoothed}, {\it does not intersect} the sphere associated to $C$.
	  Sum the Taylor coefficients to the pre-existent ones.

	}
	{\bf else}

	{\leftskip 3 true cm
	  {\bf do $B_1 \in {\cal C}(B)$}

	  {\leftskip 3.5 true cm
	    {\bf if $d_{B_1C} \geq \max\{2\epsilon_{B_1},\delta\cdot r_{B_1}\}
	    + r_{B_1}+r_C$ then}
	    
	    {\leftskip 4 true cm
	      {\bf convert} the multipole coefficients of box $B_1$ 
	      to Taylor coefficients of box $C$ with eq. (\ref{multay}),
	      because it is {\it well separated} from $C$ and the
	      the sphere where the field generated by the masses in $B_1$ is
	      {\it smoothed}, {\it does not intersect} the sphere associated to
	      $C$. Sum the Taylor coeff. to the pre-existent ones.

	    }
	    {\bf else}

	    {\leftskip 4 true cm
	      {\bf put} $B_1$ into the collection ${\cal X}(C)$.

	    }
	    {\bf end if}

	  }
	  {\bf end do}

	}
	{\bf end if}

      }
      {\bf end do}

    }
    {\bf else}  [the box $C$ is {\it terminal}]

    {\leftskip 2 true cm
      {\bf let ${\cal A} = {\cal X}(M(C))$}        
	  
      {\bf do while ${\cal A}$ not empty}
	  
      {\leftskip 2.5 true cm
	{\bf do $B \in \cal A$}
		 
	{\leftskip 3 true cm
	  {\bf if $B\in{\cal T}$ then} [the box $B$ is {\it terminal}]
	  
	  {\leftskip 3.5 true cm
	    {\bf eliminate $B$ from the set $\cal A$}
	      
	    {\bf do $i=1,n_C$}

	    {\leftskip 4 true cm
	      {\bf do $j=1,n_B$}

	      {\leftskip 4.5 true cm
		{\bf Sum} directly (i.e. without any expansion) to the grav. field
		on the particle $i$ that due to the
		particle $j$ taking into account the {\it grav. smoothing}
	     
	      } 
	      {\bf end do}

	    } 
	    {\bf end do}
      
	  }
	  {\bf else} [the box $B$ isn't terminal]

	  {\leftskip 3.5 true cm
	    {\bf if $d_{BC} \geq \max\{2\epsilon_B,\delta\cdot r_B\}+r_B+r_C$
	    then} [the box $C$ is {\it well--sep.} from the box $B$ and it is
	    outside its smoothing sphere]

	    {\leftskip 4 true cm
	      {\bf eliminate $B$ from the set $\cal A$}
		  
	      {\bf Sum} to the {\it Taylor coefficients} of the box $C$ those
	      obtained transforming the {\it multipole
	      coefficients} of the box $B$ by means of (\ref{multay}).
	      
	    }
	    {\bf end if}

	  }
	  {\bf end if}
	  
	}
	{\bf end do}
	
	{\bf let ${\cal A} = {\cal C}({\cal A})$} [Now indicate with $\cal A$
	the collection of all the children boxes of each box in the precedent
	set $\cal A$. This means that we are descending the tree to the next
	level]
	
      }
      {\bf end do}
	
      {\bf do $i=1,n_C$}

      {\leftskip 2.5 true cm
	{\bf Calculate} the local expansion of the grav. field in
	the position of the particle $i$, using the (\ref{tay1}) and the
	coefficients pertinent to the box $C$, summing to the accelerations
	calculated up to now.

      }
      {\bf end do}

    }
    {\bf end if}

  }
  {\bf end do}

}
{\bf end do}

}
}
}

Note that we have simplified the way forces on the particles in terminal
boxes are evaluated. In Greengard's adaptive algorithm this is made by means
of complicated passages and classifications of boxes into many several
collections that are computationally expensive to build up.

In our opinion this complication is unnecessary, because when one has to
consider a terminal box for wich one has the long--range component of the
potential in terms of Taylor coefficients (translated from those of its parent
box), one has only to calculate the short--range forces on the $n$ particles
(with $n<s$) inside the terminal box, due
to a certain set of near boxes and this can be done in the most
efficient way by means of the same kind of passages that in the
{\it tree--code} are used to evaluate the force on a single particle.

Suppose we have to evaluate forces on $n_C$ particles in the terminal box
$C$. When we deal with a non--terminal box $B$ and this box is {\it not}
well--separated from $C$, then it will be subdivided considering
its children boxes and the subdivision is recursively repeated until we reach
either terminal or well-separated boxes.
The contribution due to terminal boxes will be
calculated {\it directly}, that is summing particle--particle interactions.
The contibution due to
well--separated boxes will be evaluated converting their multipole expansion
coefficients into Taylor ones, summing them to the pre--existent cofficients
of the box $C$ and then, in a following passage, using these
coefficients and the Taylor expansion to evaluate gravitational forces at the
points occupied by the particle in $C$. This is done in the
last statements of the above description (from the `do while ...' forward).
\section{The matrices of coefficients}
\label{coeff}
Defining $A_j^k\equiv (-1)^j \left[(j-k)!(j+k)!\right]^{-1/2}$, we have:
\begin{eqnarray}
S^{(1)}_{j,k,n,m}&\equiv& B_{k,n}^mA_n^mA_j^k/A^{m-k}_{j+n}\\
S^{(2)}_{j,k,n,m}&\equiv& C_m^{k-m}A_n^mA_{j-n}^{k-m}/A^k_j\\
S^{(3)}_{j,k,n,m}&\equiv& D_{n-j,m-k}^mA_{n-j}^{m-k}A_j^k/A^m_n
\end{eqnarray}
where
\begin{eqnarray}
B_{k,n}^m&\equiv& (-1)^n\left\{      
\begin{array}{lr}
(-1)^{\min\{|m|,|k|\}}& \mbox{if ~~$m\cdot k>0$}\\
1&\mbox{otherwise}
\end{array} \right.\\
C_r^s&\equiv& \left\{      
\begin{array}{lr}
(-1)^{\min\{|r|,|s|\}}& \mbox{if ~~$r\cdot s<0$}\\
1&\mbox{otherwise}
\end{array} \right.\\
D_{n,m}^s&\equiv& (-1)^n\left\{      
\begin{array}{lr}
(-1)^m& \mbox{if ~~$m\cdot s<0$}\\
(-1)^{s-m}& \mbox{if ~$m\cdot s>0$ and $|s|<|m|$}\\
1&\mbox{otherwise}
\end{array} \right.
\end{eqnarray}


\end{document}